\documentclass[runningheads]{llncs}
\usepackage[pdftex]{graphicx} 
\usepackage{float} 
\usepackage{subfigure} 
\usepackage{stfloats}
\usepackage{tabularx}
\usepackage{longtable}
\usepackage{booktabs}
\usepackage{url}
\usepackage{multirow}

\DeclareGraphicsExtensions{.pdf,.jpeg,.png,.eps}

\begin{document}

\title{Do Weibo platform experts perform better at predicting stock market?}

\author{Ziyuan Ma\inst{1}\and
Conor Ryan\inst{2}\and
Jim Buckley\inst{2}\and
Muslim Chochlov\inst{2}
}

\authorrunning{Ma, Ziyuan et al.}

\institute{University \emph{of} Limerick, Department of Computer Science and Information Systems, Limerick, Ireland \\
\email{curlyjuanjuan@foxmail.com}\\
\and
University \emph{of} Limerick, LERO, Limerick, Ireland
\\
\email{\{Conor.Ryan,Jim.Buckley\}@ul.ie}
\email{Muslim.Chochlov}@lero.ie}

\maketitle

\begin{abstract}
Sentiment analysis can be used for stock market prediction. However, existing research has not studied the impact of a user's financial background on sentiment-based forecasting of the stock market using artificial neural networks. In this work, a novel combination of neural networks is used for the assessment of sentiment-based stock market prediction, based on the financial background of the population that generated the sentiment. The state-of-the-art language processing model  Bidirectional Encoder Representations from Transformers (BERT) is used to classify the sentiment and a Long-Short Term Memory (LSTM) model is used for time-series based stock market prediction. For evaluation, the Weibo social networking platform is used as a sentiment data collection source. Weibo users (and their comments respectively) are divided into Authorized Financial Advisor (AFA) and Unauthorized Financial Advisor (UFA) groups according to their background information, as collected by Weibo. The Hong Kong Hang Seng index is used to extract historical stock market change data. The results indicate that stock market prediction learned from the AFA group users is 39.67\% more precise than that learned from the UFA group users and shows the highest accuracy (87\%) when compared to existing approaches.
\end{abstract}


\section{Introduction}

There is a wide range of factors affecting the stock market: national policies, exchange rates, epidemics, amongst others \cite{french1987expected}.

Recently, several sentiment-based stock market prediction approaches have relied on social media comments and short texts (for example, 'tweets') towards stock market prediction \cite{10.1371/journal.pone.0180723,sprenger2014tweets,bollen2011twitter,mittal2012stock}. The existing approaches, however, do not differentiate sentiments according to the users' financial background.

Weibo is the largest social media platform in China that allows for qualification certificates to be provided by users as evidence of their professional background. This allows for more reliable user grouping into an Authorized Financial Advisor (AFA) group and an Unauthorized Financial Advisor (UFA) group (unauthorized and without certificates). However, to the best of our knowledge, there is currently no comparative study on the financial background of users on the Weibo platform towards the accuracy of stock market prediction. Such a study could be used towards the improvement of existing sentiment-based stock market prediction systems: for example, higher weights could be given to sentiments published by a certain user group (if known for its better prediction accuracy). The research question of this paper, therefore, is: \textbf{How do sentiments of AFA users compare to UFA users in terms of stock market prediction accuracy?}

To address this question, this work extracts stock market related comments of AFA and UFA users on the Weibo platform, uses these comments with a novel combination of neural networks for stock market prediction, and compares whether there is a difference between AFA and UFA users' comments in predicting the stock market. The main contributions of this paper can be summarized as follows:

\begin{itemize}

\item A novel combination of natural language processing neural network (BERT) and LSTM neural network towards stock market prediction;
\item Implementation of this prediction model (along with data of this experiment) that is publicly available on Github\footnote{\url{https://github.com/majuanjuan/Doexpertsperformbetter}};
\item Resulting from the first contribution, the finding that financial experts (AFA group) are 39.67\% more accurate in their stock market predictions than UFA group.
\end{itemize}

The remaining chapters of this paper are as follows. \emph{Section Two} discusses related work in sentiment analysis and development of stock market forecasting models. \emph{Section Three} explains the structure of the stock market prediction system used in this work, as well as the functions and implementation details of each module in the system. \emph{Section Four} presents the empirical design of the study and \emph{Section Five} describes the results of the empirical study. \emph{Section Six} summarizes threats to the validity of the experiment and \emph{Section Seven} draws conclusions and suggests possible future work.

\section{Related work} 

Brown and Cliff \cite{brown2004investor} have proposed the possibility of using collective sentiment indicators to predict the stock market, opening up a new research direction. For example, \cite{bollen2011twitter,mittal2012stock} used comments on the stock market and then categorize them by sentiment. Based on these results, a causal relationship between the movement of the stock market and comment sentiment was derived. More recently, Chu et al. proposed a wavelet \cite{chu2016nonlinear} method to identify the causal relationship between stock market returns and investor sentiment\cite{jin2016has}. These (social media sentiments and time series based stock prediction) are the basis of this work.

\subsection{Sentiment analysis using neural networks}

In the field of natural language processing, the methods employed in sentiment analysis problems can be divided into statistics-based sentiment dictionary methods and machine learning methods \cite{peng2017a}. 

In machine learning approaches, texts are converted into vectors, and then the classification of a vector is obtained \cite{rao2014building}. Language models such as word2vec\cite{mikolov2013efficient}, given texts, can produce word embeddings (word representations as numeric vectors) for further use in downstream tasks such as classification. However, the word2vec model cannot efficiently handle synonyms and polysemy. The BERT model solves the above-mentioned problems of the word2vec model \cite{devlin2018bert}. BERT uses transformers for representation, supplemented by a new masked language model technology to achieve bidirectional information acquisition, and can be fine-tuned for application in a wide range of natural language downstream tasks \cite{devlin2018bert}. 

\subsection{Machine learning in stock market prediction}
Traditional machine learning models such as K-Nearest Neighbors (KNN) and Support Vector Machines (SVM) were applied to stock market prediction. Wuhtrich et al. \cite{inproceedings} used several KNN to classify the stock trends by analyzing financial newspaper, the best accuracy rate was 53\%. In another research \cite{mittermayer2006newscats}, SVM was used by analyzing collected news data from PRNewswire (company-specific press releases). They showed that the average accuracy rate was 82\%. Zhenkun et al. studied stock market prediction based on Weibo comments' sentiments using SVM \cite{zhou2016can} and reported prediction accuracy of 67\%. Zhao et al. \cite{zhao2020sentiment} investigated the prediction of financial assets (housing prices, stocks, gold, etc.), created a financial dictionary for sentiment analysis, and then used this with SVM for stock market prediction. 

Kaastra and Boyd have found that deep learning models can capture rules in complex financial scenarios \cite{KAASTRA1996215}. Amongst these, LSTM seems to show a better ability to deal with long time series data \cite{hochreiter1997lstm}. Zhuge et al. used LSTM as a regression model to analyze \cite{zhuge2017lstm} stocks. The experiment used sentiment data collected from Eastmoney, stock data, and Shanghai Composite Index. This experiment used mean square error (MSE) for evaluation, the value in this experiment is in the range from 0.000152 to 0.000513, which suggests that models with sentiment analysis can perform better than those only input technical (e.g., price and volume) stock data. Recently, Jiawei and Murata \cite{jiawei2019stock} used word2vec and LSTM with a set of sentiment data to analyze the movement of the Shanghai Composite Stock Index. This research achieved an accuracy of 66\%. 

Our study uses BERT for sentiment analysis combined with LSTM for time series analysis.

\section{Stock market prediction system}
\begin{figure}[!th]
\centering
\includegraphics[scale = 0.4]{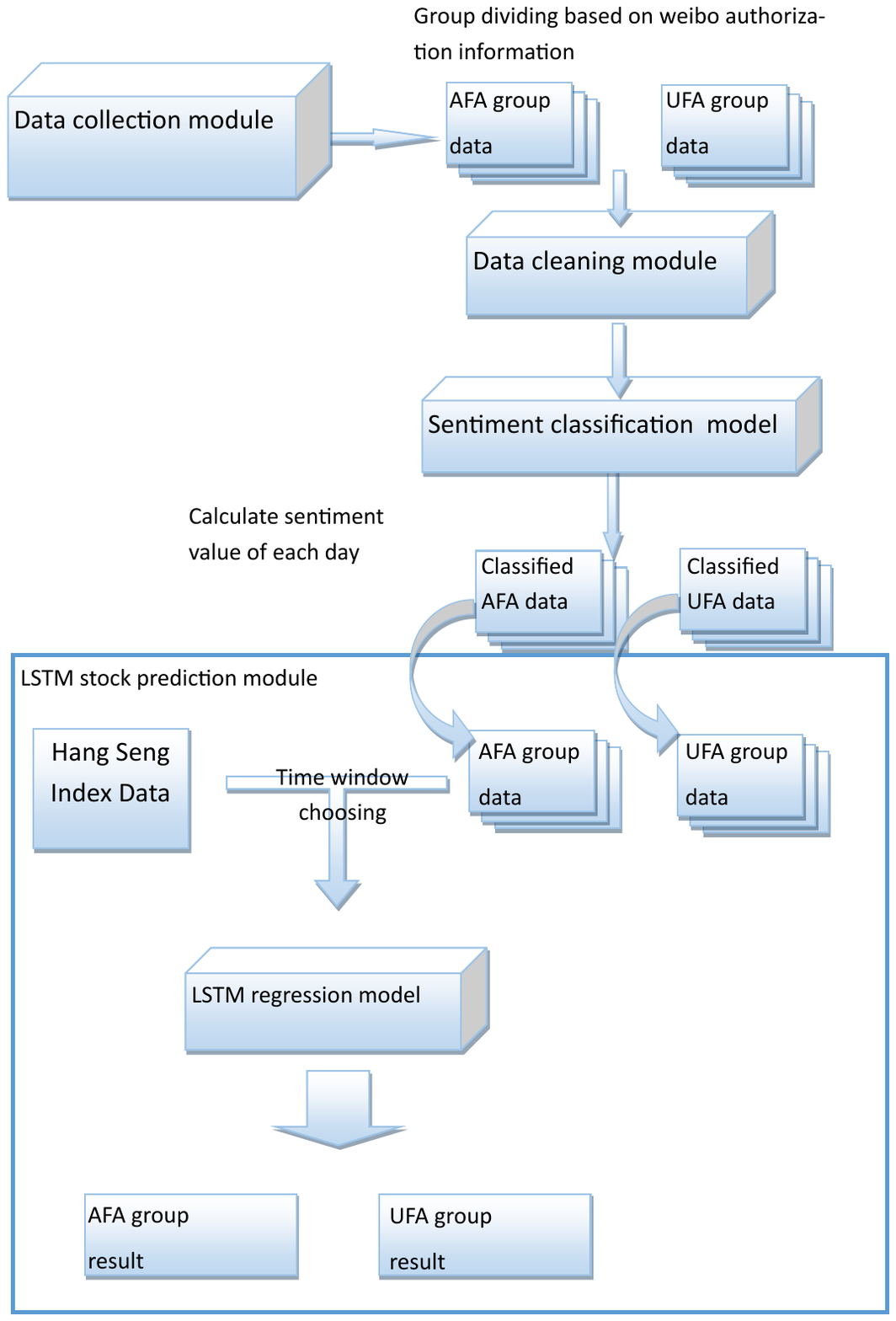}
\caption{Stock market prediction system structure}
\label{System structure}
\end{figure}

Figure~\ref{System structure} describes the structure of our stock market prediction system. First, a data collection module is used to collect Weibo text data, then the data is processed using a data cleaning module. This data is passed to the sentiment classification module for analysis and finally is used by the LSTM-based regression module for stock market price prediction. The system is our development, and the sentiment analyzing module and stock predicting module is developed based on the existing project \cite{tracy-talent,rguthrie3}. The code of the prediction system used in this work is publicly available (see Introduction section).

\subsection{Data collection module}

The data collection module collects raw Weibo data by crawling the Weibo platform (see Figure~\ref{System structure}) and collecting historical stock data from Investing.com\footnote{\url{https://www.investing.com/}}. It is a financial news website that includes daily information on the stock markets of all countries in the world and provides historical data download functions. The data set collected in this module include:

{\bfseries Weibo data}: Weibo text data includes text, the user id, the release time, the number of comments, the number of reposts, and the number of likes. During the data collection process, Weibo text messages containing the keyword ``Hang Seng Index'' (case-insensitive) in the text field were selected to extract sentiment tendency towards this index. The time range was set to January 1, 2018 and going to December 31, 2019. There were 7,169 posts, total, in the AFA group and 164,760 posts, total, in the UFA group.

{\bfseries Stock data}: The stock market data for the Hang Seng Index was downloaded starting on January 1, 2018 and going to December 31, 2019. Daily data includes the date, closing price, opening price, highest price, lowest price, total transaction amount, and the rate of change compared to the previous day.

\subsection{Data cleaning module}
The data cleaning module cleans the raw data collected by the data collection module. Specifically, this module includes text clean-up (such as special characters that cannot be processed are deleted) and duplicate removal.

\subsection{Sentiment analysis module}
\label{Sentiment analysis module}
The sentiment analysis module is used to classify the Weibo data. The structure of this module applied a fully connected layer outside of the BERT model to process the output sentence vectors from the BERT model, this method is also suggested in BERT \cite{devlin2018bert} and has been used in existing researches \cite{karimi2020adversarial,chi2021audio}. The fully connected layer's size is 768 to fit the size of the output vector. The fine-tuning step used a set of labeled data that was collected from the Weibo platform. Subsequently, in the training step, the Weibo data is tagged with a sentiment label used to indicate whether the user's view of the stock market is positive(1) (expecting prices to increase) or negative(0) (expecting prices to decrease).

\subsection{LSTM stock prediction module}
The structure of the stock predicting module uses two layers of LSTM cells, the structure of the LSTM cell follows the design in \cite{gers1999learning}. LSTM model can handle longer time series and is suitable for use in learning scenarios with strong time dependence such as the stock market \cite{chen2015lstm}. 

The stock market prediction module has two parts. The first part calculates the daily average sentiment value for both two user groups. It then calculates the causal coefficient value based on the daily sentiment value combined with stock market data based on a time-window sliding value. In this work, \textbf{T} is used to represent this time-window value. 

The second part is used to offset the calculated daily sentiment value according to the original date, according to \textbf{T}. The module then converts it into a feature and sends it to a two-layer LSTM network, the parameters can be found in \ref{Implementation details}. In this work, two input features of the LSTM network are historical stock opening price and daily sentiment value, shifted by the value of \textbf{T}. The output of this module is a prediction for the opening price for the next day.

\subsection{Implementation details}
\label{Implementation details}
The data collection module is developed based on the Weibo-search \cite{dataabc}: Weibo platform's data crawler. 

The BERT pre-trained model (its Chinese version) loaded by the module comes from the official BERT project \cite{google-research}. This original BERT model used here includes 110M parameters and 12 layers. The pre-training process was done by Google, using Wikipedia and BookCorpus data. Fine-tuning dataset of this model comes from 'weibo\_senti\_100k' data set in SophonPlus \cite{sophonplus} and can be accessed and downloaded publicly. The SophonPlus is a widely used project for the Chinese language and already has 3,000 stars on Github (a measure of a project's popularity). This data set contains a total of 10,000 Weibo text comments in the Chinese language that have been tagged with a sentiment label. The length of each Weibo text is within 140 characters, including the plain text, emojis, and other special characters. The ratio of positive examples to negative examples here is 1:1. The BERT hyper-parameters are as follows: batch size is 64, the learning rate is 2e-5, the epoch is 3, and the warm-up proportion is 0.1. Sun \cite{sun2019fine} found that these parameters can lead to a good classification with BERT. The output size of Bert for each Weibo sentence is 768 and after processed by the fully connected layer, the vector size reshapes to 2. 

The LSTM stock predicting module includes an LSTM time series prediction layer and a linear regression output layer. In the model, the hidden size is 128, layer number is 2, the dropout rate is 0.001, and batch size is 64. The input features are pairs of previous opening price and previous sentiment value. The output of this module is the predicted opening price of the next day. The data set includes 602 rows of data from January 2018 to December 2019, 60$\%$ of data is used for training and 40$\%$ for testing. Due to this experiment was not looking to tune hyperparameters and thus validation set was excessive here.

\section{Methodology} 

To explore how AFA users’ predictions on the stock market compare to UFA users, this study collects comments from these user groups, performs sentiment analysis on these comments, and uses the two sets of sentiments to predict the Hang Seng Index separately, calculating their prediction accuracy.

\subsection{User grouping method}
Liang's research \cite{guo2014event} uses the Naive Bayes method to analyze the Weibo posted by users, and divides users into celebrities, organizations/media accounts, and grassroots stars. However, the purpose of grouping in this experiment is not user attributes, but user experience background, so the method of designing grouping in this experiment is to access Weibo API and query the public user information. For an AFA user authentication, a qualification certificate is needed \cite{WeiboAuthorization}. Thus, the definition of AFA user in this experiment is that the user's account is authenticated by Weibo, and the authentication description includes the finance-related keywords. 
The users whose ``certified'' field was TRUE and authentication description included financial keywords were assigned to the AFA group. The users whose ``certified'' field was FALSE or did not include any financial keyword in the description field were assigned to the UFA group. As a result, 3,535 users were assigned to the AFA group and 17,675 to the UFA group. 


\subsection{Daily sentiment analysis method}
In order to compare the predictive ability of users in the AFA group with those in the UFA group, it is first necessary to fit the relationship between sentiment value and stock market trends to a LSTM model. For this, a daily sentiment value, based on each specific user group category was calculated. In the experiment, the average of sentiment values in a day was used to measure the daily sentiment value of a user group:
\begin{scriptsize} 
\begin{equation}
Sentiment_{d}= \frac{\sum_{i=1}^{n}  Label_{i} }{n}
\label{con:average formula}
\end{equation}
\end{scriptsize} 
In Formula~\ref{con:average formula}, $Sentiment_{d}$ stands for the sentiment value of day $d$, $n$ means the number of texts in Weibo for this date, $Label_{i}$ is the sentiment label value of a Weibo $text_{i}$ that is found on that date. The higher the sentiment value, the more positive the general feeling is because, as mentioned earlier, the label for positive is 1 and for negative is 0.


\subsection{Time window calculation method}
Zhang et al. \cite{6021163} claimed that there is a delay between prediction and stock market movements. The delay to take effect is the \textbf{T} value. The calculation process is based on the following assumption: the result of sentiment analysis has a significant correlation with the stock market after it is shifted by \textbf{T} days. And the final value of \textbf{T} is which that makes the correlation maximum. Zhang et al. studied the selection of an appropriate \textbf{T} and found the maximal correlation between prediction and stocks when \textbf{T} was 15. 

\begin{figure}[H]
\centering   
\includegraphics[scale = 0.65]{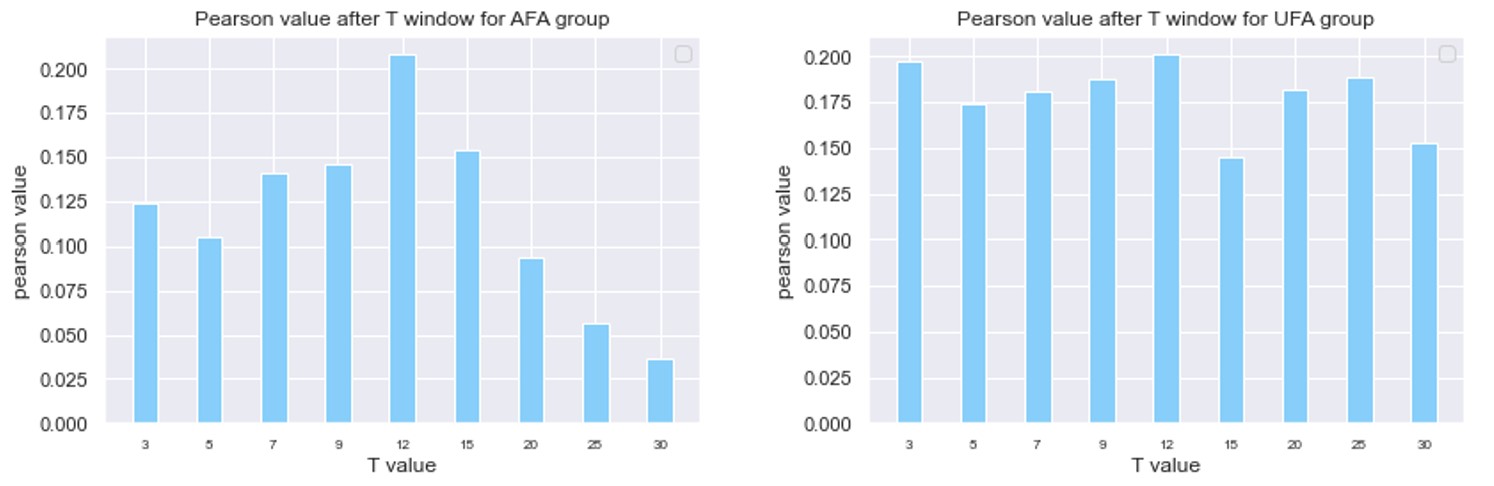}
\caption{Pearson's correlation for different T values.}
\label{Pearson with different T value}
\end{figure}

Likewise in this research, the \textbf{T} value was obtained empirically. Different time values between 3 to 30 were probed for correlation results. The calculation results after different \textbf{T}-value offsets are shown in Figure~\ref{Pearson with different T value}.
                             
As can be seen from the Figure~\ref{Pearson with different T value}, for both groups, the strongest correlation occurs when the time window value is 12 (although for the UFA group it is not as pronounced as for the AFA group). For the AFA group, Pearson's value reached above 0.200. For the UFA group, the value peaked at 0.200. Therefore, in this work, the value was set to 12 for both groups.


\subsection{Stock prediction model evaluation metrics}
As this is a regression model, the mean square error (MSE) is used to evaluate the neural network's performance. To evaluate prediction results, the confusion matrix, which is commonly used to evaluate classification models \cite{sammut2017encyclopedia}, is calculated. The definition of F1 score and precision used here can be found in Sammut's research \cite{sammut2017encyclopedia}. Finally, to compare to other research, accuracy was used, which is a fraction of all correct predictions (true positives and true negatives) out of all predictions given.

The predicted results are classified as 'rising or steady' or 'falling' based on the next day’s value. And the actual rise and fall are compared according to the trend classification results to evaluate whether the prediction is accurate. Then 'rising or steady' is assumed as a positive sample, and 'falling' is assumed as a negative sample. The actual price trend will then be compared to the predicted price trend.

\section{Results} 
\begin{figure}[H]
\centering
\includegraphics[scale = 0.4]{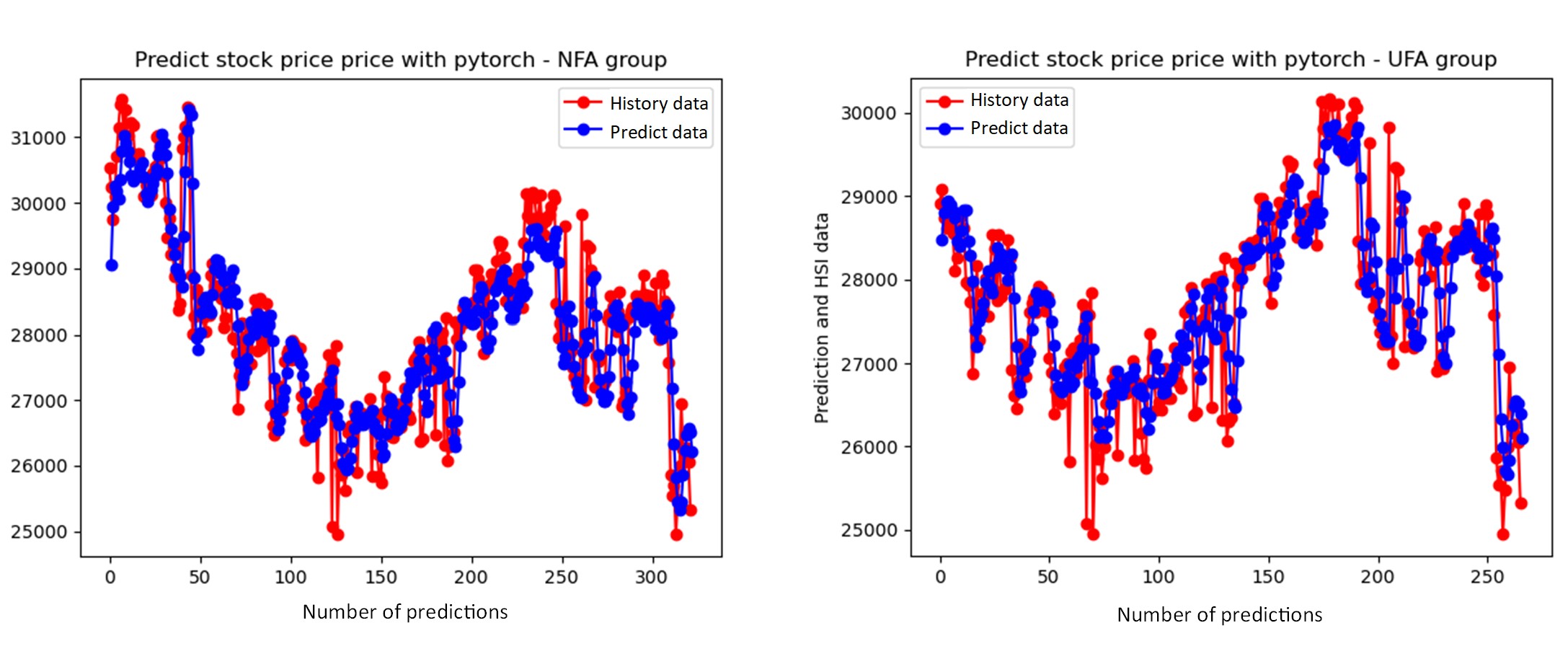}
\caption{Predict price}
\label{predict price}
\end{figure}

In this regression model, MSE is used for evaluation. For the AFA group, the MSE was 0.149 and the valid loss was 0.140. For the UFA group, the MSE was 0.076 and the valid loss was 0.211. The results of using two sets of Weibo data for prediction are shown in Figure~\ref{predict price} (y-axis in the figure shows the opening price for Hang Seng Index).

From Figure~\ref{predict price} it can be observed that the forecasts of the two groups are generally in line with the trend of the stock market, to a certain extent.

Then the results of two user groups are filled into the confusion matrix as shown in Table~\ref{tab:f1matrix}. Because the number of data in the two groups is different, the totals are not equal.

\begin{table}[htp]
\tiny
\centering
\begin{tabular}{@{}l|ll|lll@{}}
\toprule
                & \multicolumn{2}{l|}{Result matrix for AFA group} &  & \multicolumn{2}{l}{Result matrix for UFA group} \\ \midrule
                & Predicted positive     & Predicted negative     &  & Predicted positive     & Predicted negative     \\
Actual positive & 119                    & 20                    &  & 181                    & 115                     \\
Actual negative & 13                     & 103                    &  & 99                     & 172                    \\
Precision       & \multicolumn{2}{l|}{0.9015}                      &  & \multicolumn{2}{l}{0.6464}                      \\
F1 score        & \multicolumn{2}{l|}{0.2196}                      &  & \multicolumn{2}{l}{0.1571}                      \\ \bottomrule
\end{tabular}
\caption{Resultant confusion matrix and precision for AFA and UFA groups}
\label{tab:f1matrix}
\end{table}

It can be observed from Table~\ref{tab:f1matrix} that the prediction precision of the AFA group is 90.15$\%$. The prediction precision of the UFA group is 64.64$\%$. Therefore, the prediction precision of the AFA group for the stock market is 39.67$\%$ higher than that of the UFA group. The F1 score that represents the average precision and recall also shows a better score in the AFA group than the UFA group.

\begin{table}[]
\tiny
\centering
\begin{tabular}{lllr}
\hline
Author                     & Approach          & Stocks/Index predicted                                                                & Accuracy (\%) \\ \hline
Wuhtrich\_1998 {[}13{]}    & kNN               & Hang Seng Index                                                                       & 53            \\
Mittermayer\_2006 {[}14{]} & SVM               & S\&P 500 stocks                                                                       & 82            \\
Zhenkun\_2016 {[}15{]}     & SVM               & Shanghai Composite Index                                                              & 67            \\
Jiawei\_2019 {[}20{]}      & word2vec + LSTM   & \begin{tabular}[c]{@{}l@{}}Shanghai Composite Index\\ and other 6 stocks\end{tabular} & 66            \\ \hline
\multirow{2}{*}{This work} & BERT + LSTM (AFA) & Hang Seng Index                                                                       & \textbf{87}   \\
                           & BERT + LSTM (UFA) & Hang Seng Index                                                                       & 62            \\ \hline
\end{tabular}
\caption{Comparing accuracy of this work to other reported stock prediction approaches}
\label{tab:comparison}
\end{table}

When comparing the accuracy of the other existing stock prediction models (using accuracy results reported in their papers) (see Table~\ref{tab:comparison}), the model used in this work with the AFA group's sentiment-based prediction demonstrates the best accuracy results (87\%). 

\section{Threats to validity}
In the prediction result, the MSE of the AFA group model is higher than that of the UFA group's. This may be due to the difference in the size of the data set \cite{linjordet2019impact}. The data volume of the UFA group is 20 times that of the AFA group, so the model might learn more to fit the price better than in the case of the AFA group. Besides, in future work, oversampling to the AFA group can be used for balancing the scale of two data sets.

In the user grouping method, the authentication is checked by applying a financial keyword filter, users with authenticated financial keywords will be grouped into the AFA group, UFA group includes both authenticated with no financial background and non-authenticated users. Meanwhile, the classification of users is based on the authentication of Weibo: there may exist users with professional backgrounds who have not applied for certification, which leads to inaccurate grouping. Likewise, personal authentication may not be as stringent as government or enterprise authentication.

Finally, grouping rising and steady prediction results into a single group can make the comparison to other reported results inaccurate.

\section{Conclusions and Future Work}
The goal of this work was to study whether there is a difference between financial professionals (AFA group) and ordinary people (UFA group) on the Weibo social networking platform in predicting the stock market using a combination of state-of-the-art sentiment classification BERT and LSTM for stock prediction. In the study, Weibo was used as the source of sentiment extraction, and the historical data related to the Hang Seng Index was extracted. 

In this study, it was found that the AFA group performs better at predicting the stock market than the UFA group: the precision of the AFA group is 39.67$\%$ higher than that of the UFA group's. Hence, it seems that AFA posts are a much stronger foundation on which to base market predictions. The further filtering of authorization to enterprise and government-level authorization may further impact these strength-of-authorization findings and in future work, we will refine our analysis accordingly. Of course, the context of these findings should also be considered, where advisors with government authorization in a Chinese context may be considered to have heightened authority and thus may serve to make those posts have a self-fulfilling prophecy. This may have impacted our findings. and so, all three categories (personal, enterprise, and government) will be considered in isolation in our future work.

\section*{Acknowledgement}
This work was supported with the financial support of the Science Foundation Ireland grant 13/RC/2094\_2.



%

\bibliographystyle{IEEEtran}

\bibliography{references_myphd}

\begin{thebibliography}{10}
\providecommand{\url}[1]{#1}
\csname url@samestyle\endcsname
\providecommand{\newblock}{\relax}
\providecommand{\bibinfo}[2]{#2}
\providecommand{\BIBentrySTDinterwordspacing}{\spaceskip=0pt\relax}
\providecommand{\BIBentryALTinterwordstretchfactor}{4}
\providecommand{\BIBentryALTinterwordspacing}{\spaceskip=\fontdimen2\font plus
\BIBentryALTinterwordstretchfactor\fontdimen3\font minus
  \fontdimen4\font\relax}
\providecommand{\BIBforeignlanguage}[2]{{%
\expandafter\ifx\csname l@#1\endcsname\relax
\typeout{** WARNING: IEEEtran.bst: No hyphenation pattern has been}%
\typeout{** loaded for the language `#1'. Using the pattern for}%
\typeout{** the default language instead.}%
\else
\language=\csname l@#1\endcsname
\fi
#2}}
\providecommand{\BIBdecl}{\relax}
\BIBdecl

\bibitem{french1987expected}
K.~R. French, G.~W. Schwert, and R.~F. Stambaugh, ``Expected stock returns and
  volatility,'' \emph{Journal of Financial Economics}, vol.~19, no.~1, pp.
  3--29, 1987.

\bibitem{10.1371/journal.pone.0180723}
\BIBentryALTinterwordspacing
Y.~Xu, Z.~Liu, J.~Zhao, and C.~Su, ``Weibo sentiments and stock return: A
  time-frequency view,'' \emph{PLOS ONE}, vol.~12, no.~7, pp. 1--21, 07 2017.
  [Online]. Available: \url{https://doi.org/10.1371/journal.pone.0180723}
\BIBentrySTDinterwordspacing

\bibitem{sprenger2014tweets}
T.~O. Sprenger, A.~Tumasjan, P.~G. Sandner, and I.~M. Welpe, ``Tweets and
  trades: The information content of stock microblogs,'' \emph{European
  Financial Management}, vol.~20, no.~5, pp. 926--957, 2014.

\bibitem{bollen2011twitter}
J.~Bollen, H.~Mao, and X.~Zeng, ``Twitter mood predicts the stock market,''
  \emph{Journal of computational science}, vol.~2, no.~1, pp. 1--8, 2011.

\bibitem{mittal2012stock}
A.~Mittal and A.~Goel, ``Stock prediction using twitter sentiment analysis,''
  \emph{Standford University, CS229 (2011 http://cs229. stanford.
  edu/proj2011/GoelMittal-StockMarketPredictionUsingTwitterSentimentAnalysis.
  pdf)}, vol.~15, 2012.

\bibitem{brown2004investor}
G.~W. Brown and M.~T. Cliff, ``Investor sentiment and the near-term stock
  market,'' \emph{Journal of empirical finance}, vol.~11, no.~1, pp. 1--27,
  2004.

\bibitem{chu2016nonlinear}
X.~Chu, C.~Wu, and J.~Qiu, ``A nonlinear granger causality test between stock
  returns and investor sentiment for chinese stock market: a wavelet-based
  approach,'' \emph{Applied Economics}, vol.~48, no.~21, pp. 1915--1924, 2016.

\bibitem{jin2016has}
X.~Jin, D.~Shen, and W.~Zhang, ``Has microblogging changed stock market
  behavior? evidence from china,'' \emph{Physica A: Statistical Mechanics and
  its Applications}, vol. 452, pp. 151--156, 2016.

\bibitem{peng2017a}
H.~Peng, E.~Cambria, and A.~Hussain, ``A review of sentiment analysis research
  in chinese language,'' \emph{Cognitive Computation}, vol.~9, no.~4, pp.
  423--435, 2017.

\bibitem{rao2014building}
Y.~Rao, J.~Lei, L.~Wenyin, Q.~Li, and M.~Chen, ``Building emotional dictionary
  for sentiment analysis of online news,'' \emph{World Wide Web}, vol.~17,
  no.~4, pp. 723--742, 2014.

\bibitem{mikolov2013efficient}
T.~Mikolov, K.~Chen, G.~S. Corrado, and J.~Dean, ``Efficient estimation of word
  representations in vector space,'' 2013.

\bibitem{devlin2018bert}
J.~Devlin, M.-W. Chang, K.~Lee, and K.~Toutanova, ``Bert: Pre-training of deep
  bidirectional transformers for language understanding,'' \emph{arXiv preprint
  arXiv:1810.04805}, 2018.

\bibitem{inproceedings}
B.~Wuthrich, V.~Cho, S.~Leung, D.~Permunetilleke, K.~Sankaran, and J.~Zhang,
  ``Daily stock market forecast from textual web data,'' vol.~3, 11 1998, pp.
  2720 -- 2725 vol.3.

\bibitem{mittermayer2006newscats}
M.~A. Mittermayer and G.~Knolmayer, ``Newscats: A news categorization and
  trading system,'' pp. 1002--1007, 2006.

\bibitem{zhou2016can}
Z.~Zhou, J.~Zhao, and K.~Xu, ``Can online emotions predict the stock market in
  china?'' in \emph{International conference on web information systems
  engineering}.\hskip 1em plus 0.5em minus 0.4em\relax Springer, 2016, pp.
  328--342.

\bibitem{zhao2020sentiment}
W.~Zhao, F.~Wu, Z.~Fu, Z.~Wang, and X.~Zhang, ``Sentiment analysis on weibo
  platform for stock prediction,'' in \emph{International Conference on
  Artificial Intelligence and Security}.\hskip 1em plus 0.5em minus 0.4em\relax
  Springer, 2020, pp. 323--333.

\bibitem{KAASTRA1996215}
I.~Kaastra and M.~Boyd, ``Designing a neural network for forecasting financial
  and economic time series,'' \emph{Neurocomputing}, vol.~10, no.~3, pp. 215 --
  236, 1996, financial Applications, Part II.

\bibitem{hochreiter1997lstm}
S.~Hochreiter and J.~Schmidhuber, ``Lstm can solve hard long time lag
  problems,'' in \emph{Advances in neural information processing systems},
  1997, pp. 473--479.

\bibitem{zhuge2017lstm}
Q.~Zhuge, L.~Xu, and G.~Zhang, ``Lstm neural network with emotional analysis
  for prediction of stock price.'' \emph{Engineering letters}, vol.~25, no.~2,
  2017.

\bibitem{jiawei2019stock}
X.~Jiawei and T.~Murata, ``Stock market trend prediction with sentiment
  analysis based on lstm neural network,'' in \emph{Proceedings of the
  International MultiConference of Engineers and Computer Scientists}, 2019,
  pp. 13--15.

\bibitem{tracy-talent}
\BIBentryALTinterwordspacing
Tracy-Talent, ``tracy-talent/curriculum.'' [Online]. Available:
  \url{https://github.com/tracy-talent/curriculum/tree/master/Data
  Mining/sentiment}
\BIBentrySTDinterwordspacing

\bibitem{rguthrie3}
\BIBentryALTinterwordspacing
rguthrie3, ``rguthrie3/deeplearningfornlpinpytorch.'' [Online]. Available:
  \url{https://github.com/rguthrie3/DeepLearningForNLPInPytorch}
\BIBentrySTDinterwordspacing

\bibitem{karimi2020adversarial}
A.~Karimi, L.~Rossi, A.~Prati, and K.~Full, ``Adversarial training for
  aspect-based sentiment analysis with bert,'' \emph{arXiv preprint
  arXiv:2001.11316}, 2020.

\bibitem{chi2021audio}
P.-H. Chi, P.-H. Chung, T.-H. Wu, C.-C. Hsieh, Y.-H. Chen, S.-W. Li, and H.-y.
  Lee, ``Audio albert: A lite bert for self-supervised learning of audio
  representation,'' in \emph{2021 IEEE Spoken Language Technology Workshop
  (SLT)}.\hskip 1em plus 0.5em minus 0.4em\relax IEEE, 2021, pp. 344--350.

\bibitem{gers1999learning}
F.~A. Gers, J.~Schmidhuber, and F.~Cummins, ``Learning to forget: Continual
  prediction with lstm,'' 1999.

\bibitem{chen2015lstm}
K.~Chen, Y.~Zhou, and F.~Dai, ``A lstm-based method for stock returns
  prediction: A case study of china stock market,'' in \emph{2015 IEEE
  international conference on big data (big data)}.\hskip 1em plus 0.5em minus
  0.4em\relax IEEE, 2015, pp. 2823--2824.

\bibitem{dataabc}
\BIBentryALTinterwordspacing
Dataabc, ``dataabc/weibo-search.'' [Online]. Available:
  \url{https://github.com/dataabc/weibo-search}
\BIBentrySTDinterwordspacing

\bibitem{google-research}
\BIBentryALTinterwordspacing
Google-Research, ``google-research/bert.'' [Online]. Available:
  \url{https://github.com/google-research/bert}
\BIBentrySTDinterwordspacing

\bibitem{sophonplus}
\BIBentryALTinterwordspacing
SophonPlus, ``Sophonplus/chinesenlpcorpus.'' [Online]. Available:
  \url{https://github.com/SophonPlus/ChineseNlpCorpus/tree/master/datasets/}
\BIBentrySTDinterwordspacing

\bibitem{sun2019fine}
C.~Sun, X.~Qiu, Y.~Xu, and X.~Huang, ``How to fine-tune bert for text
  classification?'' in \emph{China National Conference on Chinese Computational
  Linguistics}.\hskip 1em plus 0.5em minus 0.4em\relax Springer, 2019, pp.
  194--206.

\bibitem{guo2014event}
L.~Guo, W.~Wang, S.~Cheng, and X.~Que, ``Event-based user classification in
  weibo media,'' \emph{The Scientific World Journal}, vol. 2014, 2014.

\bibitem{WeiboAuthorization}
\BIBentryALTinterwordspacing
W.~platform, ``Weibo customer service.'' [Online]. Available:
  \url{https://kefu.weibo.com/faqdetail?id=20831}
\BIBentrySTDinterwordspacing

\bibitem{6021163}
{Kaihui Zhang}, {Lei Li}, {Peng Li}, and {Wenda Teng}, ``Stock trend
  forecasting method based on sentiment analysis and system similarity model,''
  in \emph{Proceedings of 2011 6th International Forum on Strategic
  Technology}, vol.~2, 2011, pp. 890--894.

\bibitem{sammut2017encyclopedia}
C.~Sammut and G.~I. Webb, \emph{Encyclopedia of machine learning and data
  mining}.\hskip 1em plus 0.5em minus 0.4em\relax Springer, 2017.

\bibitem{linjordet2019impact}
T.~Linjordet and K.~Balog, ``Impact of training dataset size on neural answer
  selection models,'' in \emph{European Conference on Information
  Retrieval}.\hskip 1em plus 0.5em minus 0.4em\relax Springer, 2019, pp.
  828--835.

\end{thebibliography}

\end{document}